# Astro2020 Science White Paper

# GRBs as Probes of the Early Universe with *TSO*

**Thematic Areas:** ☐Planetary Systems ☐Star and Planet Formation
☒Formation and Evolution of Compact Objects ☒Cosmology and Fundamental Physics
☒Stars and Stellar Evolution ☐Resolved Stellar Populations and their Environments
☒Galaxy Evolution ☐Multi-Messenger Astronomy and Astrophysics

**Principal Author:**
Name: Nial Tanvir
Institution: Leicester University, UK
Email: nrt3@leicester.ac.uk
Phone: +44 7980 136499

**Co-authors:** Jonathan Grindlay[1], Edo Berger[1], Brian Metzger[2], Suvi Gezari[3], Zeljko Ivezic[4], Jacob Jencson[5], Mansi Kasliwal[5], Alexander Kutyrev[6], Chelsea Macleod[1], Gary Melnick[1], Bill Purcell[7], George Rieke[8], Yue Shen[9], Michael Wood Vasey[10]

[1]Center for Astrophysics | Harvard and Smithsonian, USA [2]Columbia University, USA, [3]University of Maryland, USA, [4]University of Washington, USA, [5]Caltech, USA, [6]NASA/GSFC [7]Ball Aerospace, USA, [8]University of Arizona, USA, [9]University of Illinois, USA, [10]University of Pittsburgh, USA

**Abstract:** Long gamma-ray bursts (GRBs) are the most luminous known electromagnetic radiation sources in the Universe for the ~3 – 300 sec of their prompt flashes (isotropic X/$\gamma$ -ray luminosities up to ~$10^{53}$ erg $s^{-1}$). Their afterglows have first day rest-frame UV/optical absolute magnitudes AB ~ -30 – -23. This luminous continuum nUV-nIR back-light provides the ultimate probe of the SFR(z) back to the first Pop III–II.5 stars, expected to be massive and GRB progenitors. GRB afterglow spectra in the first ~1–3 hours will directly measure their host galaxy ionization fraction $x_i$ vs. z in the Epoch of Reionization (EOR), tracing the growth of structure. Only 28% of *Swift* GRBs have measured redshifts due to limited followup at R, J <21. Some ~25% of GRBs are optically dark due to dust absorption in their host galaxies, but those with low NH in their X-ray spectra are likely at z >7. Current 8-10m telescopes and coming ELTs cannot pursue optically dark GRBs promptly, nor can *JWST* or WFIRST slew within ~0.5-1 days of a GRB. The Time-domain Spectroscopic Observatory (*TSO*) is a proposed Probe-class 1.3m telescope at L2, with imaging and spectroscopy (R = 200, 1800) in 4 bands (0.3 - 5$\mu$m) and rapid slew capability to 90% of sky. *TSO* would *finally* utilize z > 6 - 12 GRBs as the most direct probe of the SFR(z), EOR(z), and possibly the first direct detection of the core collapse of the very first (Pop III) stars.

## Present state of GRB science at z >6

Thanks to the immense luminosity of the long gamma-ray bursts, both their prompt and afterglow emission, they are detectable in principle to redshifts *z* ~20 (e.g. Racusin+08). Long-duration GRBs arise from massive star core collapse and thus trace star formation and star-forming galaxies through cosmic history. Furthermore, their afterglows have simple power-law spectral energy distributions making them ideal back-lights for absorption line studies of their host ISM, intervening galaxies and the IGM. The near-infrared is particularly important, since at redshifts z >7 the opacity of the intergalactic medium blocks all light in the observed optical band.

Considerable progress has been made in the past ~14 years thanks to the *Swift* satellite, which has detected more than 1200 long-GRBs. However, only ~30% have measured redshifts, illustrating the limits of the small number and restricted geographical distribution of 8m telescopes world-wide, which are usually required to perform the photometry to find the optical/nIR counterpart and then perform spectroscopy to measure a redshift. A significant fraction of X-ray afterglows are optically dark, in many cases due primarily to dust absorption in their host galaxy but for others due to high redshift. For both scenarios, the range of available nIR spectrographs is even more limited. Nonetheless, thanks to *Swift* we have identified nine GRBs with z ≥6 (e.g. Tanvir et al. 2018; also Fig. 1), and studies of nearly complete sub-samples suggest that at least twice that number have likely been discovered but not recognized as being high-z events due to limited follow-up (Perley+2016).

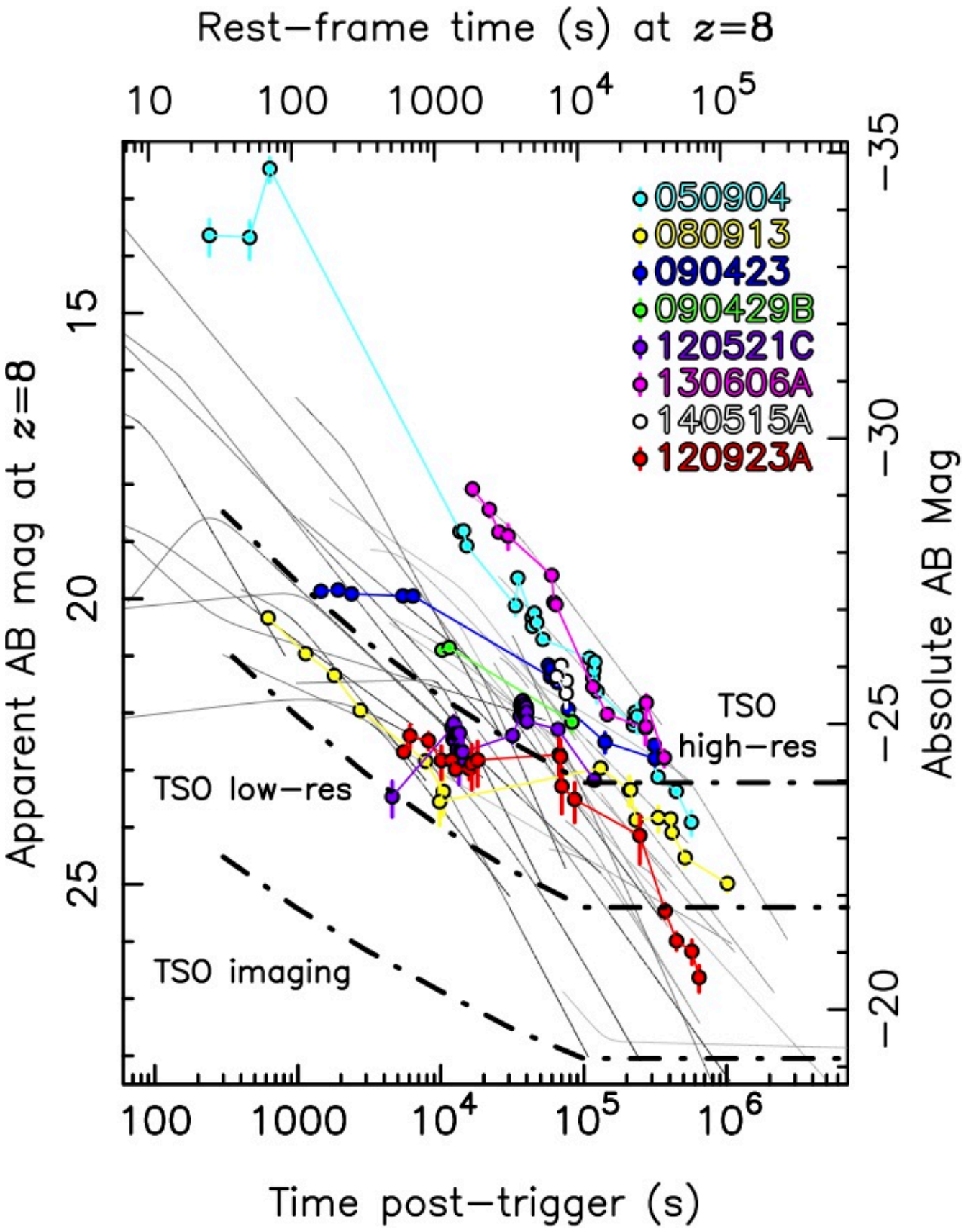

***Figure 1.*** The light curves of 8 GRB afterglows with z >6 (colored points), together with the light curves of well-sampled lower redshift GRBs (gray lines), *rescaled to a common redshift of z=8.* The 10$\sigma$ limits for imaging and for low- and high-resolution near-IR spectroscopy with a space-based telescope derived for the *TSO* mission concept are indicated. These capabilities would provide redshifts for all similar GRBs, and detailed spectroscopic diagnostics for a high proportion of the brighter afterglows.

## Star formation rate, SFR(z), at z > 4

Mapping star formation over cosmic time is a long-standing goal of cosmology. Traditional techniques generally rely on assessing the star formation in galaxies detected to a photometric limit in some waveband, and correcting for star formation and galaxies missed due to this selection bias. At *z* ≥ 6 the galaxy luminosity function appears to steepen considerably so that most star formation occurs in very faint galaxies, below the limits of surveys such as the Hubble Ultra-Deep Field, not going far enough into the IR. Thus the corrections become large, and sensitive to assumptions such as the minimum galaxy luminosity. *JWST* will make major advances in measuring this, but since each GRB represents the death of a massive star, counting GRBs as a function of redshift with *TSO* reflects the evolution of global star formation, particularly of the populations of high mass stars likely responsible for reionization. Beyond $z$ ~ 4, the apparent preference of long-GRBs for sub-solar metallicity environments becomes a minor concern, and we expect the GRB rate to be directly proportional to SFR (Greiner+2015). Furthermore, the ratio of detected to undetected hosts in deep imaging directly constrains the fraction of SFR(z) occurring in galaxies below a given detection limit (Tanvir+2012; Trenti+2012).

***Abundances and conditions in early galaxies***

In addition to pinpointing their hosts, high signal-to-noise spectroscopy of the GRB afterglows, possible with *TSO* (Fig. 2), not only provides redshifts, but also allows precision determination of abundances and physical conditions in the interstellar media of their hosts (e.g. Thoene+2013; Hartoog+2015). Thus the chemical imprints of even earlier generations of stars, including population III stars, can be studied directly with deep early spectroscopy. This will provide more detailed and precise gas-phase abundance and IGM determinations than possible from emission line diagnostics for faint galaxies. From the mid-2020s, the presumed availability of ~25-30m class ground-based telescopes can obtain (weather and <1 day response time permitting) even higher resolution nIR spectra for *pre-selected (by TSO) z >7* GRBs up to the highest redshifts.

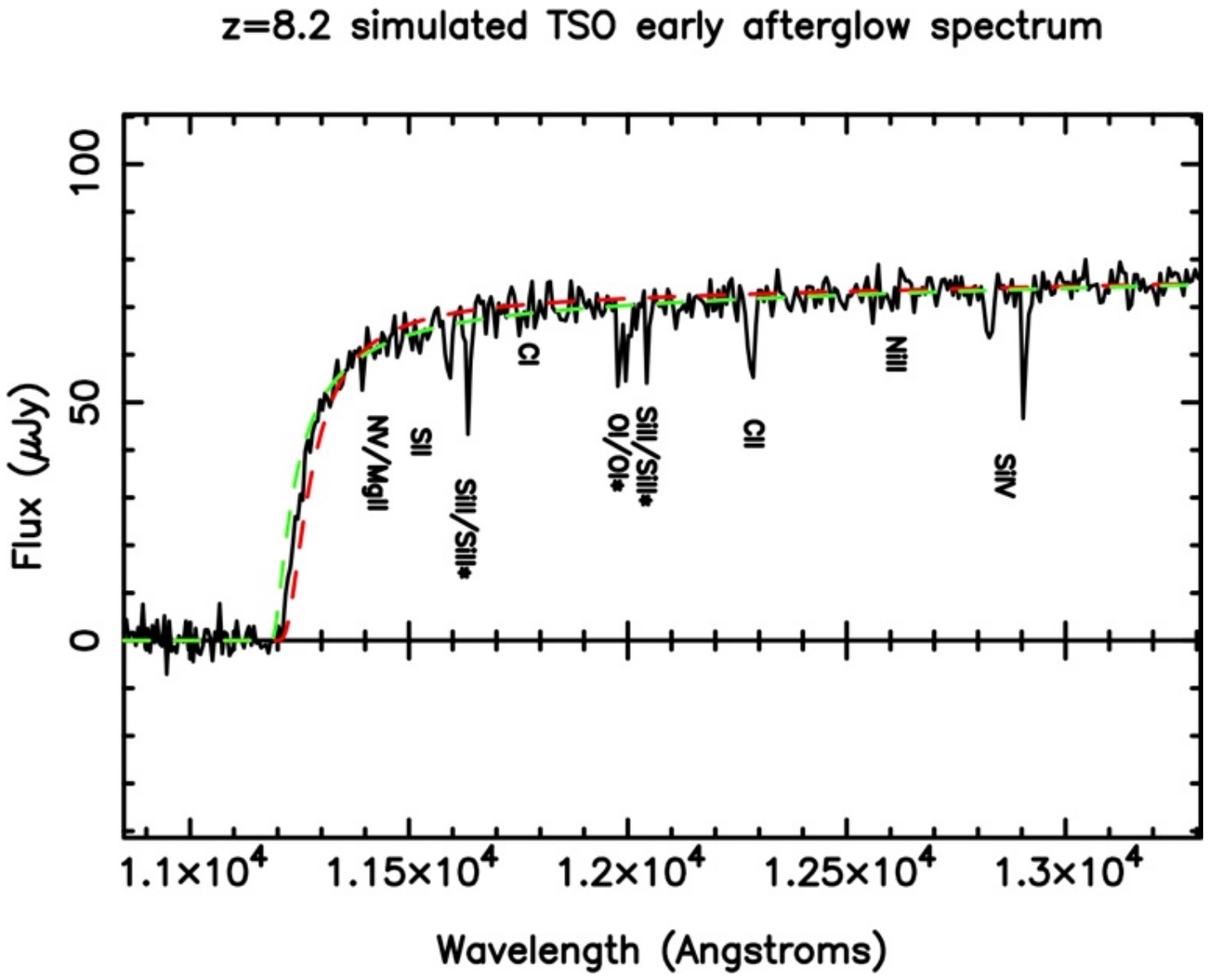

***Figure 2.*** A simulated spectrum for the high-resolution spectroscopic mode of *TSO*, based on the luminosity of the known z=8.2 GRB 090423 as it would appear if observed within the first few hours post-burst. The signal-to-noise is sufficient to identify numerous metal species, and to distinguish between Lyman-$\alpha$ absorption due to the host (**red** model) and intergalactic medium (**green** model).

***The timeline and sources of reionization***

Spectroscopy also allows us to characterize the damping wing of Lyman-$\alpha$, which for sufficiently high S/N can be decomposed to estimate the neutral hydrogen column in both the host galaxy and surrounding intergalactic medium (Fig. 2). The latter can be used to monitor the neutral fraction and hence the progress of reionization at the redshift of the burst (e.g. McQuinn+08); while the former sets a lower limit on the opacity of the sight line to the escape of ionizing radiation. The average escape fraction is very hard to measure directly, but is a key quantity in determining whether or not early generations of massive stars could have been the primary driver of reionization, with required values of >~10% generally being cited. Current samples of GRB sight lines provide a strong statistical limit on the escape fraction at z<5 (see Fig. 3; Tanvir+13), which is uncomfortably below at <~1%, but the sample at z>6 needs to be enlarged to apply this test in the era of reionization.

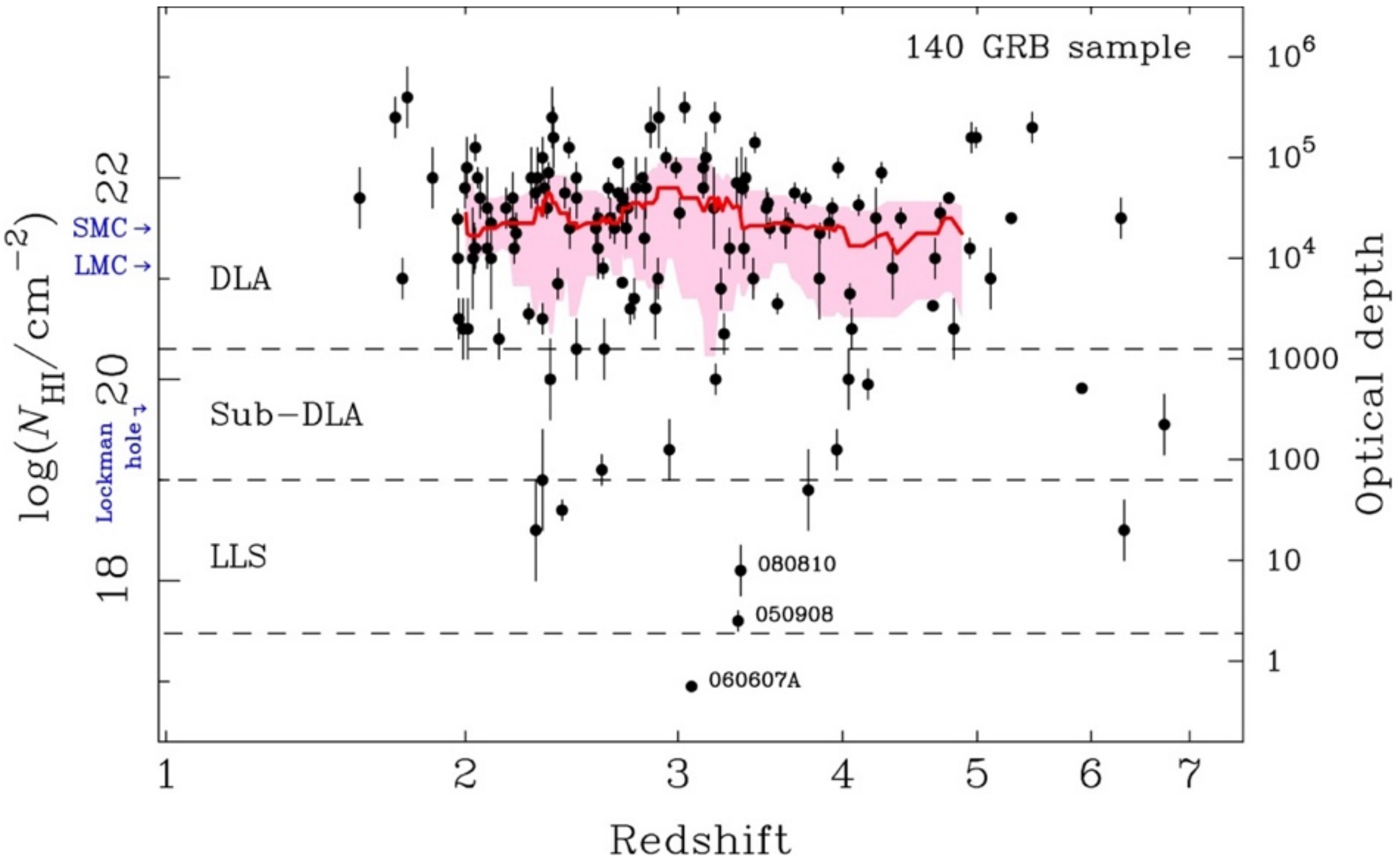

***Figure 3.*** A plot of the host HI column density on the lines of sight to all existing GRBs for which it has been measured. The inferred optical depth to radiation below the Ly-limit is indicated on the right-hand axis, showing that all but two sight-lines are effectively opaque. If these sight-lines are representative of those to massive stars which dominate extreme UV production, this implies a very low escape fraction of ionizing radiation of <1% (Tanvir+19). Although the sample of bursts with HI measures at z>6 is small, there is little indication of a substantial drop, which presents a problem for the hypothesis that the reionization of the Universe was brought about primarily by star-light. The need for a substantially larger sample at z >6, which *TSO* would provide, is obvious.

***Population III (and II.5) stars***

The discovery of examples of the 1st and 2nd generations of metal-free stars would be a landmark in extragalactic science. Theoretical considerations lead us to expect that their mass function was considerably more top-heavy than that of later generations, and so a higher fraction have likely ended their lives rapidly in core-collapse events. It has been predicted that these populations would give rise to collapsar-like events, similar to those producing regular long-GRBs, but likely of even longer intrinsic duration, and made longer still by time dilation at their high redshift (Meszaros and Rees 2010). Such events offer the *best prospect for direct detection of individual Pop III stars*, but are likely rare and will require sensitive and instantaneous $4\pi$ imaging X-ray surveys to discover and locate them, such as the *$4\pi$ X-ray Imaging Observatory (4piXIO*; see Grindlay et al, *4piXIO* WP).

***Requirements for Early Universe Exploration with GRBs***

First and foremost, is the requirement for a telescope in space for both rapid-response imaging and spectroscopy to all GRBs unless there has already been an optical counterpart identified. The requirement for a space telescope with optical–nIR (at least; but see below for *TSO* bands) imaging and spectroscopy is self-evident: even 8-10m telescopes on the ground typically only reach limiting magnitudes in JHK~21 for moderate exposures (limited by the fading time-scale) due to the high backgrounds from atmospheric OH emission. Deeper limits are possible, of course, but only with very long exposures. GRB afterglow fluxes decrease over elapsed time, T, as $\sim T^{-(1-1.3)}$ whereas limiting detection fluxes only decrease (in the low photon counting domain) as $T^{-0.5}$. So the afterglow is fading faster (except for the relatively brief extended prompt emission and initial afterglow plateau phases) and so decreased sky and telescope backgrounds are essential. This requires *TSO* in space, with no OH background, be a *cold telescope* (passive radiatively cooled to a constant 110K) to achieve zodiacal light backgrounds and enable rapid high-z GRB identification from image dropouts. High S/N immediate spectroscopy can then be done to enable the high-z GRB science goals, with the brightest initial post-burst exposures (see Fig. 2).

*TSO*, with its rapid slew capability and 4-band (nUV to mid-IR) imaging and spectroscopy (see Grindlay+ *TSO* Mission Science WP), will respond to prompt GRB alerts from Swift/BAT (if still operational) or (more likely) higher sensitivity and spatial resolution future GRB missions. *TSO* would compute and execute the safe (Sun avoidance) slew to the GRB. A first image would confirm the target field and nearby (to the GRB position) bright stars (in all 4 bands) to set exposure times scaled by the hard X-ray fluence of the discovery GRB (proportional to afterglow flux). Science images then identify IR counterparts for GRBs at z >7 by their optical dropout in the 4 band images taken simultaneously by *TSO* (see *TSO* WP). This can be done automatically on board by immediate cross-correlation or image subtraction of the images in bands 1 (0.30 – 0.73$\mu$m) and 2 (0.73 – 1.39$\mu$m) to the blue and red, respectively, of their dividing wavelength 0.73$\mu$m (chosen to be the Lyman break for z = 7) to identify the optical dropout object closest to the < 10 - 300" GRB position. *TSO* would then do an autonomous offset pointing onto the object for spectroscopy with the IFU on *TSO* for an immediate R = 200 spectrum of the GRB and surrounding 10 x 10 x 0.5" pixels to characterize the host galaxy. If the GRB has AB <23 (cf. Fig 1), subsequent long-slit spectra at R = 1800 (otherwise IFU) have increasing exposure times to achieve comparable depth despite the fading of the afterglow.

***What sample size of GRBs at z > 6 is needed and possible with TSO?***

To fully exploit z > 6 – 10 GRBs to explore and measure the Early Universe, the *total number of GRBs detected at z > 6 with high S/N spectra must increase significantly. TSO*, as a dedicated rapid-response nIR telescope in space, will do this given its continuous coverage and field of regard (fraction of sky instantly observable) that is ~80% from a Geosynch orbit or ~93% (for ≥30$^{o}$ Sun avoidance) from L2 with (much) lower particle backgrounds. Slews and settling times for <1" pointing are ≤10min for any point on the sky. In the *TSO* era (late 2020's?), *4piXIO* (see WP) could provide 3 – 200 keV full-sky/full-time GRB locations (<10") with 3 – 50 keV flux limits >1.3X below the 15-50 keV flux limit of *Swift/BAT*. Given the GRB flat photon index over the 15 – 50 keV band, *4piXIO* detects ~1.3X the BAT GRBs in this band; over the same 15 – 150 keV band, the *4piXIO* and BAT sensitivities are comparable. Finally. the $4\pi$ (4piXIO) vs. 1.4sr (BAT) FoV comparison means that instead of the *Swift/BAT* 10 year average GRB rate of 92/yr, *4piXIO* would detect 1455/yr over the same band and thus an average *TSO* GRB rate ~1/6 hours. This means that only GRBs with optical dropouts will be followed up with deep high resolution spectra. The detection rate is difficult to estimate, given the limited GRB followup and sensitivity: a 300sec image exposure with *TSO* reaches AB ~24.5, or 3.5 magnitudes fainter than the deepest (V = 21) *Swift* UVOT lower limits provided for ~90% of all GRB triggers. For TSO, All GRBs will be first observed with a 600sec R = 200 IFU spectrum so that a uniform SFR(z) study is obtained.

An extreme lower limit on the likely sample of z >6 GRBs would be that of the 364 redshifts (from 28% of GRBs detected thus far by *Swift/BAT*), 9 are at z >6, for a high-z fraction 2.5% of redshifts. Since nIR photometry coverage was only available for a small fraction (~10%; only the multi-band *GROND* telescope routinely gave JHK upper limits) of these optical/nIR searches, this ~2.5% high-z yield is a strong lower limit. Since *TSO* would have nearly 100% imaging and IFU spectra for its GRB samples, it is reasonable to apply the >2.5% high-z redshift fraction to the total GRB rate (per year) for an extreme lower limit of 1455*0.025 = 36 GRBs/yr with z >6. Since this does not allow for the >3mag sensitivity increase of every *TSO* image and IFU spectrum vs. the vast majority of ground-based image and spectra searches thus far, it is reasonable to estimate that this sensitivity increase and logN-logS provide another factor of 3 on the z >6 numbers for an approximate lower limit of ~100 GRBs/year. This is a significant sample of sight-lines that will make possible the first studies of clumpiness in the EOR and IGM. These could be followed up with deep SKA imaging to provide high resolution 21cm maps along these sight-lines. Since these static sightlines are now available for very deep spectroscopy, JWST and WFIRST can followup with more detailed searches for discrete sources of ionization: QSOs, extended star formation sites, and shocks from galaxy mergers. *TSO* (and full-sky, full-time GRB detection) can and will "light the way".